\documentclass[twocolumn,showpacs,amsmath,amssymb,showkeys,10pt,nofootinbib,superscriptaddress]{revtex4-1}
\usepackage{epsfig}
\usepackage{latexsym}
\usepackage{graphicx}
\usepackage{amsfonts}
\usepackage{amsmath}
\usepackage{todonotes}
\usepackage{ulem}
\usepackage{xcolor}
\usepackage{wasysym}
\usepackage{bm}
\usepackage{array}
\usepackage{enumitem}

\begin{document}

\title{Testing gravity on cosmological scales with cosmic shear, cosmic microwave background anisotropies, and redshift-space distortions}
\author{Agn\`es Fert\'e}
\affiliation{Institute for Astronomy, University of Edinburgh, Blackford Hill, Edinburgh EH9 3HJ, UK}
\author{Donnacha Kirk}
\affiliation{Department of Physics and Astronomy, University College London, Gower Street, London WC1E 6BT, UK}
\author{Andrew R.~Liddle}
\affiliation{Institute for Astronomy, University of Edinburgh, Blackford Hill, Edinburgh EH9 3HJ, UK}
\affiliation{Instituto de Astrof\'{\i}sica e Ci\^{e}ncias do Espa\c{c}o,
Faculdade de Ci\^{e}ncias, Edif\'{\i}cio C8, Campo Grande, 1769-016 Lisboa, Portugal}
\affiliation{Perimeter Institute for Theoretical Physics, 31 Caroline Street North, Waterloo, Ontario N2J 2Y5, Canada}
\author{Joe Zuntz}
\affiliation{Institute for Astronomy, University of Edinburgh, Blackford Hill, Edinburgh EH9 3HJ, UK}

\date{\today}
\begin{abstract}
We use a range of cosmological data to constrain phenomenological modifications to general relativity on cosmological scales, through modifications to the Poisson and lensing equations. We include cosmic microwave background anisotropy measurements from the {\it Planck} satellite, cosmic shear from CFHTLenS and DES-SV, and redshift-space distortions from BOSS data release 12 and the 6dF galaxy survey. We find no evidence of departures from general relativity, with the modified gravity parameters constrained to $\Sigma_0 = 0.05_{-0.07}^{+0.05}$ and $\mu_0 =  -0.10^{+0.20}_{-0.16}$, where $\Sigma_0$ and $\mu_0$ refer to deviations from general relativity today and are defined to be zero in general relativity. We also forecast the sensitivity to those parameters of the full five-year Dark Energy Survey and of an experiment like the Large Synoptic Survey Telescope, showing a substantial expected improvement in the constraint on  $\Sigma_0$.
\end{abstract}

\maketitle

\section{Introduction}

The acceleration of the expansion of the Universe (as first confirmed by Refs.~\cite{Riess_1998} and \cite{Perlmutter_1999}) is well described by a cosmological constant $\Lambda$.
That said, the margin of error on the relevant measurements from current data is still sufficiently large that alternative mechanisms are under investigation. These include dynamical dark energy models which, like the cosmological constant, are  best thought of as new gravitating sources appearing in the stress-energy tensor, on the right-hand side of Einstein's equations.

Alternatively, the same observations could be described by a new understanding of the laws of gravity. These would appear as modifications of the Einstein tensor on the left-hand side of Einstein's equations, codifying how gravitating sources cause space-time to curve.
Various modifications to general relativity (GR) have been proposed (such as $f(R)$ gravity, the Horndeski models and many more) that could explain the acceleration of the expansion --- see \textit{e.g.}\ Ref.~\cite{Clifton_2012} for a review of these theories and their cosmological implications. 
A phenomenological approach has also been investigated, as introduced in Ref.~\cite{Amendola_2008}, where two parameters $\Sigma_0$ and $\mu_0$ embody the deviations to GR at the linear perturbation level. This has the advantage of being flexible and observation-led, with the downside that it is not easy to map a given parameterised deviation from GR to a particular, consistent modification of the action/Einstein equations.

Various efforts have been made to constrain deviations from GR in the framework of this phenomenological parametrisation using different cosmological observables. The most recent include cosmic microwave background (CMB) temperature and polarisation anisotropy measurements from the {\it Planck} satellite \cite{planck_demg_2015}, redshift-space distortion (RSD) measurements from BOSS DR12 \cite{Mueller_2016}, and cosmic shear measurements from CFHTLens \cite{simpson_2013} or the more recent survey KIDS \cite{Joudaki_2016} and in combination with redshift-space distortions from 2dFLenS and BOSS in Ref.~\cite{Joudaki_2017}.
In all these studies $\Lambda$CDM is consistent with the data, except for Ref.~\cite{planck_demg_2015} where a tension with general relativity was found (see Figs.~14 and 15 of that paper).
The tension is caused by a degeneracy between the amplitude of the lensing $A_{\rm L}$ and modified gravity (MG) parameters leading to deviation from GR in MG. But this tension is reduced when using CMB lensing as it does not prefer a higher $A_{\rm L}$.

In this paper, we update the CFHTLenS constraints on the MG parameters $\Sigma_0$ and $\mu_0$ (as found in Ref.~\cite{simpson_2013}) by combining its comic shear measurements with the latest RSD and CMB measurements, which contain complementary information on modified gravity parameters. We use the CMB measurements from the {\it Planck} satellite and RSD measurements from BOSS DR12 and 6dFGS. We also show the dependence of the constraints on the modelling of intrinsic alignments, which is one of the most important weak-lensing systematics. Furthermore through analysis of the effect of scale cuts we are able to include a greater range of scales than the analysis in Ref.~\cite{planck_demg_2015}. We also analyze the Dark Energy Survey Science Verification (DES-SV) weak-lensing dataset for comparison, with both analyses serving as prototypes for planned future analysis of DES lensing data in constraining modified gravity.


\section{Phenomenological parametrisation of modified gravity}
\label{sec2}

Our goal is to test the behaviour of gravity on cosmological scales. 
For this purpose we adopt a phenomenological approach as used in several previous analyses such as Refs.~\cite{bean_2010,simpson_2013,Dossett_2015,Mueller_2016,Joudaki_2016}. 
In this approach the deviations from general relativity are directly related to observables. Instead of adopting a specific modification to the Einstein--Hilbert action, we consider the deviations at the level of the equations governing the evolution of large-scale structures. 
This approach does not require us to choose a specific theory of modified gravity and is therefore generic.
Ultimately, these generic deviations from general relativity can be related to the theoretical parameters of the Effective Field Theory of dark energy developed in Ref.~\cite{Gubitosi_2013}, as was shown in Ref.~\cite{Pogosian_2016} and followed-up in Ref.~\cite{Peirone_2018}.

Using the conformal Newtonian gauge, the line element in a flat Friedmann--Lema\^{\i}tre--Robertson--Walker metric with scalar perturbations is defined by: 
\begin{equation}
ds^2 = a^2(\tau) \left[ -(1 + 2\psi) d\tau^2 + (1 - 2\phi) \delta_{ij} dx_i dx_j \right] \,,
\end{equation}
with $\psi$ and $\phi$ the scalar perturbations, corresponding to the Bardeen potentials. In this paper, we assume there is no anisotropic stress. The Einstein field equations then give the following constraint equations: 
\begin{eqnarray}
\label{eq:constraint1}
k^2 \psi = & -  4 \pi G a^2 \rho \delta \,, \\
\label{eq:constraint2}
k^2 (\psi + \phi) =  & - 8 \pi G a^2 \rho \delta \,.
\end{eqnarray}
where $\delta$ is the density contrast and $\rho$ the mean matter density. These equations imply that $\phi = \psi$ in general relativity, in absence of anisotropic stress.

Various modifications of the $\psi$ and $\phi$ potentials and their combinations have been introduced in the literature. We refer the reader to Ref.~\cite{planck_demg_2015} for a summary of these parameterizations.
In this paper, we adopt the ($\Sigma_0$,$\mu_0$) parameterization as used in Refs.~\cite{simpson_2013,Zhao_2010}. These two parameters modify the above Eqs.~(\ref{eq:constraint1}) and (\ref{eq:constraint2}) to become
 \begin{eqnarray}
\label{eq:mu}
k^2 \psi = & -  4 \pi G a^2 \left[1 + \mu(a,k) \right] \rho \delta \,, \\
\label{eq:sigma}
k^2 (\psi + \phi) =  & - 8 \pi G a^2 \left[ 1 + \Sigma(a,k)\right] \rho \delta \,.
\end{eqnarray}
The parameter $\mu_0$ modifies the gravitational potential $\psi$ felt by non-relativistic tracers such as galaxies through our RSD probe.    
The parameter $\Sigma_0$ acts on the sum of the potentials $\phi + \psi$ which is felt by the non-relativistic tracers such as the photons. Gravitational lensing from cosmic shear and lensing of the CMB is therefore sensitive to $\Sigma_0$.  

In general the $\Sigma_0$ and $\mu_0$ parameters can depend on both scale and time, giving considerable freedom for potential deviations from GR. Current data is not sufficient to deliver precise measurements across all possible scales/times. We must reduce the flexibility of our parameterised deviations to maximise our ability to produce useful constraints.
In this paper, we only consider a time dependence, in order to limit the number of degrees of freedom to constrain and therefore be able to draw conclusions on time dependence of deviations from GR. 
Furthermore, Ref.~\cite{planck_demg_2015} has shown that adding a scale dependence in their analysis did not improve the overall fit; they therefore chose to consider scale-independent parameters. As we are expecting the same level of constraining power as in their analysis, we assume scale-independence as well. 
However, the different datasets that we use in this paper are probing different scales, and so combining them to constrain scale-independent parameters can lead to artificially small constraints. Therefore a more thorough study of the MG parameters scale dependence should be undertaken, especially in the case of future experiments where the constraining power will be greater. 

\begin{figure}[t]
\begin{center}
\includegraphics[scale=0.5]{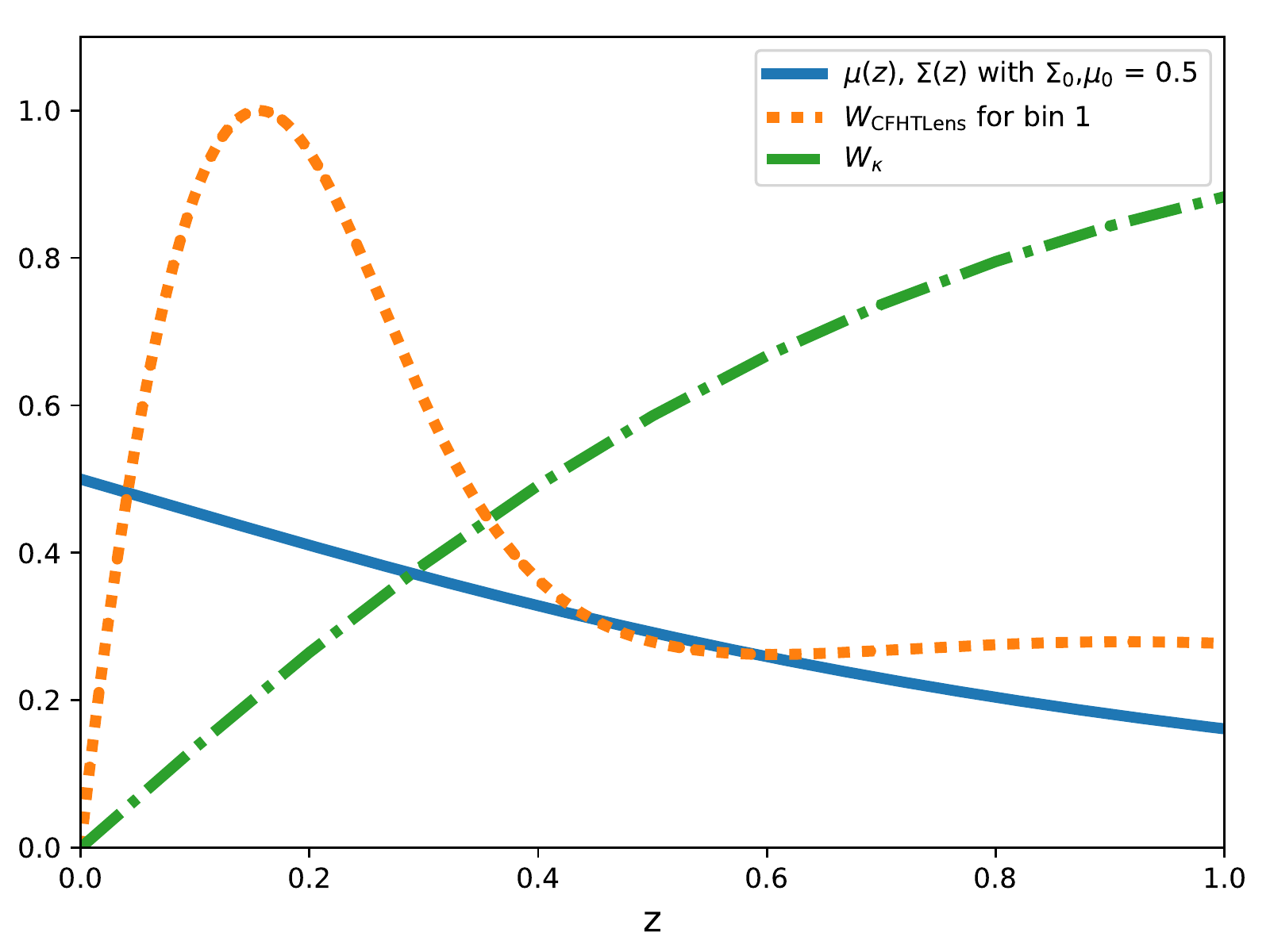}
\caption{The redshift evolution of $\Sigma_0$ and $\mu_0$ in blue for a value of 0.5 at $z = 0$. The kernels for CFHTLenS $W_{\mathrm{CFHTLenS}}$ and the CMB lensing $W_{\kappa}$ are shown in orange and green respectively, normalised by their maximum. }
\label{fig:mgparam}
\end{center}
\end{figure}

We adopt here the same time parametrisation as the one in Ref.~\cite{simpson_2013}:
\begin{equation}
X(a) = X_0 \frac{\Omega_{\Lambda}(a)}{\Omega_{\Lambda,0}},
\end{equation}
where $X$ stands for $\Sigma$ or $\mu$. The motivation for this parametrisation is that at the background level, a deviation from GR can be described by a dark energy, and such a deviation would plausibly impact the perturbation evolution proportionally to the dark energy energy density. Figure~\ref{fig:mgparam} shows this time parametrisation for $X_0 = 0.5$. Also shown are the kernels of the weak-lensing (here for the first bin of CFHTLenS) and CMB lensing observables, to show where their peak contributions lie compared to these time-dependent modified gravity parameters.

Throughout this paper we assume the background evolution matches $\Lambda$CDM. We are seeking to explain the same, observed, accelerated expansion by testing deviations from GR at the level of the perturbations rather than invoking a cosmological constant. Nevertheless if the non-GR phenomenology delivers the same background expansion, we can treat this as an  effective $\Lambda$ on the level of the background calculations. 

Clearly this phenomenological approach can only capture a rough picture of how a fundamental theory might modify observables. In particular a fundamental theory might imply specific relations between modified background and perturbation evolution. Nevertheless, this is appropriate for the `pre-discovery' phase of modified gravity theories, where the objective is to find any generic hints as to failures of GR, particularly in the absence of any single compelling candidate modified gravity theory to test directly.

We will be combining different probes sensitive to $\Sigma_0$ and $\mu_0$. This parameterization has been designed so that the $\mu_0$ parameter is predominantly constrained by non-relativistic tracers, in our case RSD. Conversely the $\Sigma_0$ parameter is primarily constrained by relativistic tracers, which we access via cosmic shear and lensing of the CMB. Both parameters are highly degenerate with standard cosmological parameters so the accurate CMB measurements from \textit{Planck} are necessary in constraining cosmology and breaking these degeneracies. Moreover, secondary CMB anisotropies due to the integrated Sachs--Wolfe effect make the CMB anisotropies on large scales sensitive to $\Sigma_0$ and $\mu_0$ as well. We will pay particular attention to the impact of systematics on the weak-lensing measurements in the context of MG studies as these must be well understood and controlled if upcoming weak-lensing projects like DES, {\it Euclid}, and LSST are to reach their full potential.

We use the parameter estimation code CosmoSIS \cite{Zuntz_2015} to sample cosmological and nuisance parameters. 
To make the theoretical predictions we use the Boltzmann code \textsc{mgcamb} \cite{Zhao_2009,Hojjati_2011} which  is available within CosmoSIS; we have modified it in order to include the ($\Sigma_0$,$\mu_0$) model. We derive the parameters constraints using \textsc{multinest} \cite{Feroz_2009,Feroz_2008,Feroz_2013}. 

The priors we adopt on $\Sigma_0$ and $\mu_0$ are flat within the range $-3$ to $3$. We marginalise over the amplitude and the spectral index of the primordial fluctuations $A_{\rm s}$ and $n_{\rm s}$, the energy density of matter and baryons $\Omega_{\rm m}$ and $\Omega_{\rm b}$, the reduced Hubble parameter $h_0 = H_0/100$, the optical depth $\tau$. When cosmic shear is included, we marginalise over the shear magnitude bias $m_i$, the photometric bias $b_i$ for redshift bin $i$ and the amplitude of the intrinsic alignment $A_{\mathrm{IA}}$. We will detail these systematics in Section \ref{sec:shear}. When the CMB is included (through the \textit{Planck} data as described later), we marginalise over the  \textit{Planck} nuisance parameter $A_{\mathrm{P}}$. 
Our priors on the standard cosmological, systematics, and nuisance parameters are shown in Table~\ref{tab:priors}. 

\begin{table}[t]
\centering
\begin{tabular}{|c |l|} 
 \hline
 \textbf{~Parameters~} & \textbf{\qquad \quad ~~Priors} \\
 \hline\hline
\multicolumn{2}{|c|}{\textbf{$\Lambda$CDM}} \\ 
\hline
$A_{\rm s}$ & flat [$0.5 \times 10^{-9};5 \times10^{-9}$]	\\
$n_{\rm s}$ & flat [0.8;1.2]					\\ 
$\Omega_{\rm m}$ & flat [0.05;0.9]			\\ 
$\Omega_{\rm b}$ & flat [0.01;0.08]			\\
$h_0$ & flat [0.5,1] 					\\
$\tau$ & flat [0.01;0.2]				\\
 \hline
\multicolumn{2}{|c|}{ \textbf{Modified Gravity}} 	\\
  \hline
 $\Sigma_0$ &flat [$-3$;3]	 \\
 $\mu_0$ & flat [$-3$;3]\\
 \hline
\multicolumn{2}{|c|}{ \textbf{Shear systematics}} \\ 
  \hline
 $m_{i}$ & gaussian $\sigma$ = 0.02 \\
 &  [$-0.1$;0.1] (CFHTLenS)	 \\
  & gaussian $\sigma$ = 0.05 \\
  & [$-0.2$;0.2] (DES-SV) \\
 $\delta z_{i}$ & gaussian $\sigma$ = 0.02 \\
 & [$-0.1$;0.1] (CFHTLenS) \\
 & gaussian $\sigma$ = 0.05 \\
 & [$-0.3$;0.3]  (DES-SV)\\
$A_{\mathrm{IA}}$ & flat [$-5$,5]\\
 \hline
\multicolumn{2}{|c|}{  \textbf{{\it Planck} nuisance parameter}} \\ 
  \hline
$A_{\mathrm{P}}$ & gaussian $\sigma$ = 0.0025  \\
& [0.9,1.1] \\
\hline
\end{tabular}
\caption{Priors on the cosmological and modified gravity parameters, along with the weak-lensing systematics and \textit{Planck} nuisance parameter for the CMB, used in the present work. }
\label{tab:priors}
\end{table}


\section{Cosmological observables, dataset and parameter estimation}

\label{sec3}

In this section we describe the cosmic shear, RSD, and CMB datasets used in this paper and present their individual constraints on $\Sigma_0$ and $\mu_0$, before combining them in the following section. 

\subsection{Cosmic shear}
\label{sec:shear}

\subsubsection{Observable, data set and systematics}

\begin{figure}[t]
\begin{center}
\includegraphics[scale=0.5]{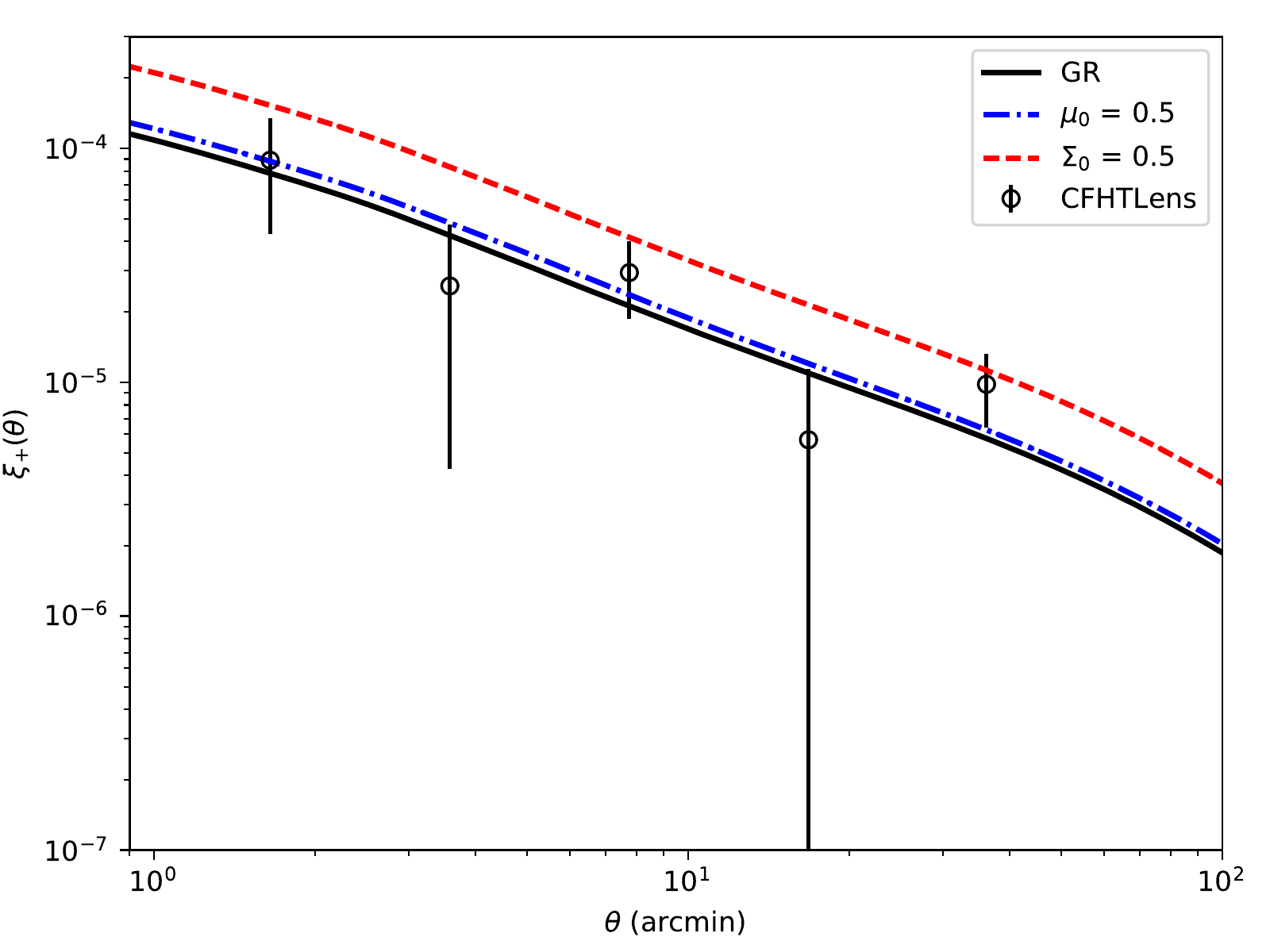}
\includegraphics[scale=0.5]{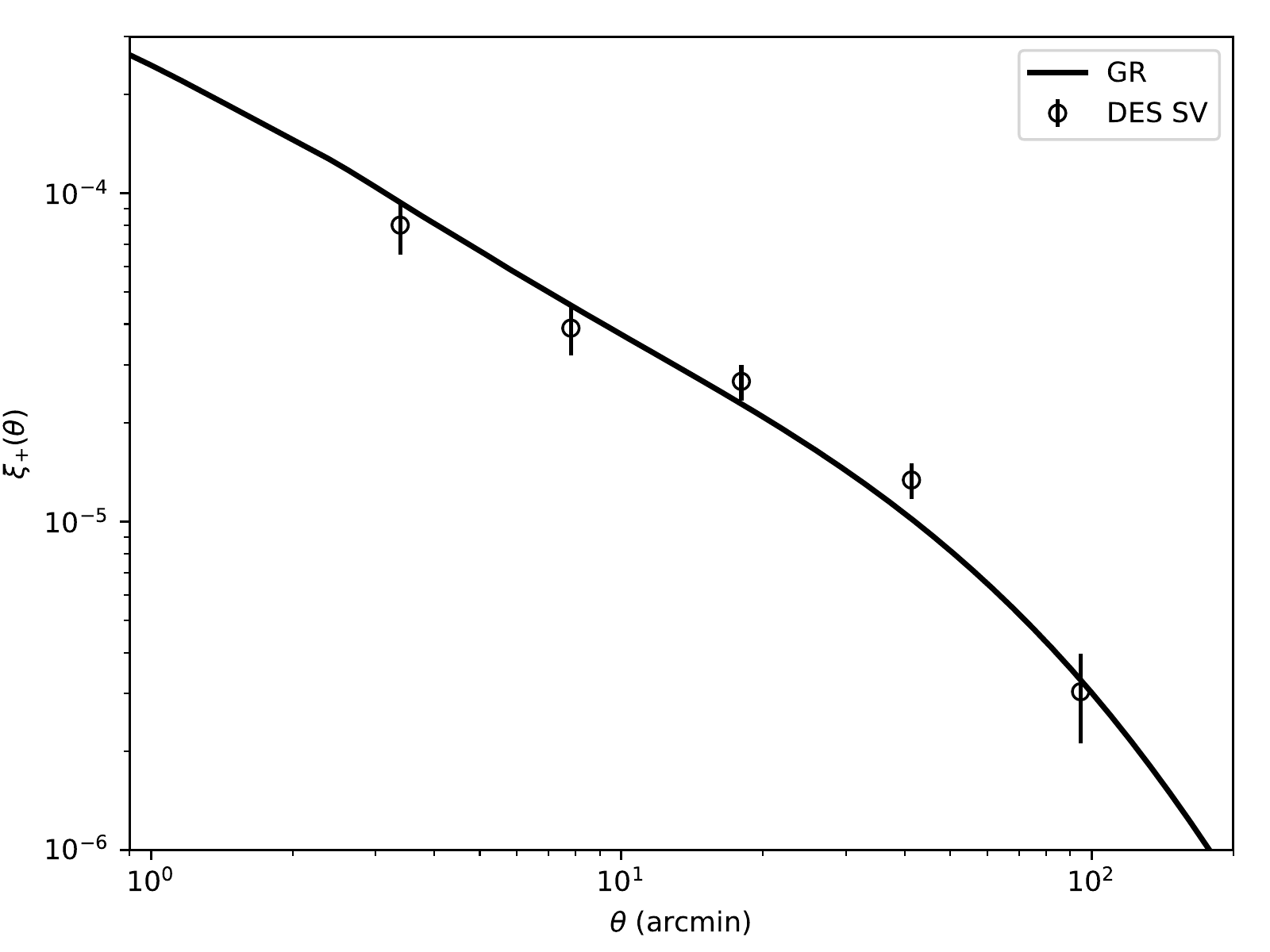}
\caption{Angular 2-point correlation function $\xi_+(\theta)$ measurements from the full CFHTLenS sample used in Ref.~\cite{Heymans_2013}  in redshift bin 3 (upper panel) and DES-SV in Ref.~\cite{shear_dessv} (lower panel). The black curve represents the theoretical $\xi_{+}(\theta)$ for the case of general relativity (i.e.\ $(\Sigma_0,\mu_0) = (0,0)$). In the upper panel, the blue dot-dashed curve shows $\xi_+(\theta)$ with $(\Sigma_0,\mu_0)=(0.5,0)$ and the red dashed curve  $(\Sigma_0,\mu_0)=(0,0.5)$.}
\label{fig:xi}
\end{center}
\end{figure}

Cosmic shear is the distortion of galaxy images through gravitational lensing by large-scale structures between the observer and the observed galaxy. 
The angular \mbox{2-point} correlation functions $\xi_{\pm}^{i,j}(\theta)$ of the cosmic shear in different redshift bins $i$ and $j$ is a powerful probe as it can trace the distribution of dark matter and the evolution of structures.
The modification of the laws of gravity alters this quantity mainly through the convergence power spectrum, which depends on the $\Sigma_0$ parameter as shown in Fig.~\ref{fig:xi}. 

Many projects are dedicated to measuring cosmic shear, such as the CFHTLenS \cite{Heymans_2013} and KIDS \cite{Hildenrandt_2016} surveys, the ongoing Dark Energy Survey (DES) \cite{DES_2005}, HSC \cite{hsc_2017} and the forthcoming {\it Euclid} satellite \cite{Euclid_2011}, LSST lensing surveys \cite{LSST_2012}, and WFIRST \cite{wfirst_2012}.
In this study we use the CFHTLenS data, one of the most precise measurements of cosmic shear to date.\footnote{For the KiDS modified gravity study, see Ref.~\cite{Joudaki_2016}.} 
This survey covered 154 square degrees of the sky in 5 optical bands. 
In the present analysis, we exploit the full CFHTLenS sample used in Ref.~\cite{Heymans_2013}, which consists of $\xi_{\pm}(\theta)$ measurements in 6 redshift bins ranging from 0.20 to 1.30 and for 5 angles ranging from 1.5 to 35 arcmin. 

For comparison, we also analyze constraints from the DES Science Verification (DES-SV) data. DES is an ongoing survey which will cover 5000 square degree of the sky in 5 near-infrared and optical bands. The DES-SV data were taken between November 2012 and February 2013, before the beginning of the DES survey proper, and cosmology results were presented in Ref.~\cite{dessv_2016}. The cosmic shear measurements were made in 3 redshift bins of 0.3 to 1.3 for angular scales ranging from 2 to 60 arcminutes. Because of the lower depth and sky coverage of the survey, its constraining power on ($\Sigma_0$,$\mu_0$) is less than CFHTLenS (as is the case for $\sigma_8$ and $\Omega_{\rm m}$ as shown in Ref.~\cite{dessv_2016}). 

Figure \ref{fig:xi} shows an example of the CFHTLenS $\xi_+$ measurements for redshift bin 3, and DES-SV measurements for redshift bins 3 (which do not match the CFHTLenS redshift bins). 

We consider three sources of systematics in weak-lensing surveys. 
First, the shape measurements of the galaxies can be subject to error which can lead to a multiplicative factor $m_i$, for the redshift bin $i$, in the observed cosmic shear spectra. We marginalise over $m_i$, with priors for CFHTLenS and DES-SV listed in Table~\ref{tab:priors}. 
Second, the photometric redshift probability distribution $n(z)$ can suffer from calibration uncertainties $\delta z_i$, in redshift bin $i$, that will introduce a shift in $n(z)$. We marginalise over $\delta z_i$ with the priors shown in Table~\ref{tab:priors}.
Another major systematic in weak-lensing observations is the galaxy intrinsic alignment (IA) \cite{Catelan_2001}. The shape of galaxies can be influenced by the gravitational environment where the galaxies formed. If neglected, this can bias cosmological parameter estimation from cosmic shear \cite{Joachimi_2011, Kirk_2012, Krause_2016}. In the present work, we use the non-linear alignment model from Refs.~\cite{Bridle_2007, Kirk_2012}, adapted from Ref.~\cite{Hirata_2004}, as used in the CFHTLenS and DES-SV analysis \cite{Heymans_2013, dessv_2016}. In this model, the IA power spectra are proportional to the non-linear matter power spectrum with a free amplitude $A_{\mathrm{IA}}$. We assume that the formation of structure takes place early enough that any deviation to GR was negligible at that time, and so do not introduce any MG related modifications to the intrinsic alignment model. We marginalise over the amplitude $A_{\mathrm{IA}}$ with the priors quoted in Table~\ref{tab:priors}.

\subsubsection{Impact of systematics}

First, we note that the choice of scale cuts in the 2-point statistics is an important issue in cosmic shear analyses. The current theoretical predictions of the matter power spectrum on small scales are indeed subject to uncertainties due to baryonic effects. To avoid this source of error, the \mbox{2-point} correlation function measurements on smaller angular scales are usually not included in current cosmological analysis. 
Furthermore, we currently do not have a model for non-linear scales in the case of the ($\Sigma(a,k)$,$\mu(a,k)$) parametrisation.

In this work, we decided to adopt the scale cuts as defined in Ref.~\cite{Heymans_2013}, where non-linear scales have been removed based on deviations of $\xi_{\pm}$ for a 7$\%$ boost or decrease of the non-linear corrections from Ref.~\cite{Smith_2003} to the matter power spectrum. This leads to minimum angular scale $\theta_{\mathrm{min}}$ = 3 arcmin for redshift bin combinations including redshift bins 1 and 2 for $\xi_+$, and $\theta_{\mathrm{min}}$ = 30(16) arcmin for redshift bin combinations including redshift bins 1, 2, 3 and 4 (bins 5 and 6) for $\xi_-$ (corresponding to discarding 92 data points over 210). 
However, the treatment of small scales in the context of constraining modified gravity is going to be important for forthcoming weak-lensing surveys. It has been for instance explored in Ref.~\cite{Casas_2017} in the case of two semi-analytical approximations to treat non-linearities for future experiments such as {\it Euclid}, SKA and DESI.

We now turn to the impact of systematics on the ($\Sigma_0$,$\mu_0$) constraints. To explore this we derive the constraints without including the effects in our theory model (and hence not marginalising over the relevant nuisance parameters). 

\begin{figure}[t]
\begin{center}
\includegraphics[scale=0.6]{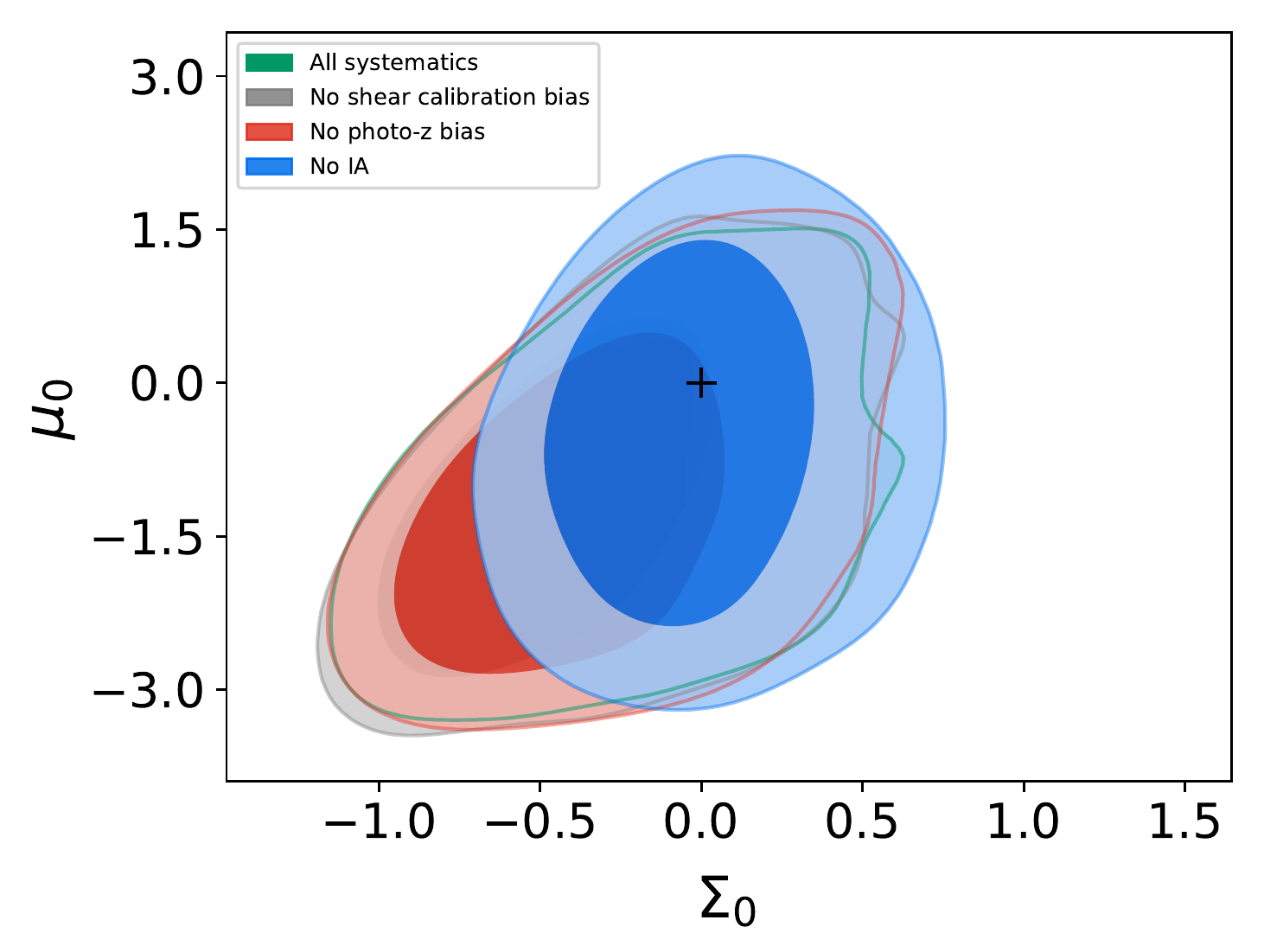}
\caption{68$\%$ and 95$\%$ confidence contours on the modified gravity parameters $\Sigma_0$ and $\mu_0$ using all the $\xi_{\pm}$ CFHTLenS measurements. Green shows marginalisation over all cosmological parameters and systematics in green. Grey ignores the presence of a shear calibration bias, red ignores the presence of photometric redshift bias, and blue ignores the presence of intrinsic alignment.}
\label{fig:syste}
\end{center}
\end{figure}

Figure~\ref{fig:syste} shows the results: we see that ignoring the shear calibration bias or the photometric redshift bias does not significantly impact the constraints on ($\Sigma_0$,$\mu_0$). On the other hand, the intrinsic alignment does shift the constraints towards higher $\Sigma_0$. 
The power from IA in the cosmic shear comes as $-A_{\mathrm{IA}}$, the amplitude of IA  multiplied by -1. And the amplitude of the IA of the full galaxy sample in the CFHTLenS analysis favors a negative $A_{\mathrm{IA}}$, although still consistent with zero. So ignoring the presence of IA in the data will favor a higher value of $\Sigma_0$ to compensate.
This study indicates the importance of IA in tests of gravitation with weak lensing. We further explore the impact of IA on the estimation of ($\Sigma_0$, $\mu_0$) in Section~\ref{sec:IA}. 

Moreover, the cosmic shear does constrain $\Sigma_0$ more than $\mu_0$. 
The constraints on MG from CFHTLenS, using only quasi-linear scales and marginalisng over the nuisance parameters, are: 
\begin{eqnarray}
\Sigma_0 & = & -0.7^{+0.5}_{-0.2} \,; \\
\mu_0 & = &  -1.6 \pm 1.0 \,.
\end{eqnarray}

\subsection{Growth of structure}

\subsubsection{Observable and data set}

The growth rate of structure can be probed by redshift-space distortions (RSD), which originate from the peculiar velocity of galaxies. 
These distortions cause anisotropies in the galaxy 2-point correlation function in redshift space \cite{Kaiser_1987}. Measured using redshift surveys such as BOSS \cite{Marin_2016} or WiggleZ \cite{Blake_2011}, the anisotropy in the clustering and its amplitude give us access to the combination of the growth function $f$ and the power spectrum amplitude $\sigma_8$. 
The parameter $\mu_0$ modifies the potential $\psi$, so the growth of the linear matter density perturbations therefore depends on $\mu_0$. RSD are the observable of choice to constrain the modified growth of structure. In our analysis, we solve for the growth equation, Eq.~(11) in Ref.~\cite{Linder_2003}, extended to include $\mu_0$. We note that the constraints from the RSD are fully independent of $\Sigma_0$. 

In the present work, we use the cosmological results presented in Ref.~\cite{alam_2016} from the DR12 galaxy sample of the Baryon Oscillation Spectroscopic Survey (BOSS), dedicated to the measurements of the Baryonic Accoustic Oscillation and RSD. This data set gives competitive measurements of the growth for mid-redshifts. 
In particular, we use the consensus values and covariances of the Hubble parameter $H(z)$, the comoving angular-diameter distance $D_m(z)$, and $f\sigma_8$ at the three redshifts: $z = 0.38$, $0.51$, and $0.61$.  
We also require a low-redshift measurement of $f\sigma_8$ to improve our constraining power, and use the 6dFGS measurement at $z = 0.067$ \cite{beutler_2012}. 

\begin{figure}[t]
\begin{center}
\includegraphics[scale=0.5]{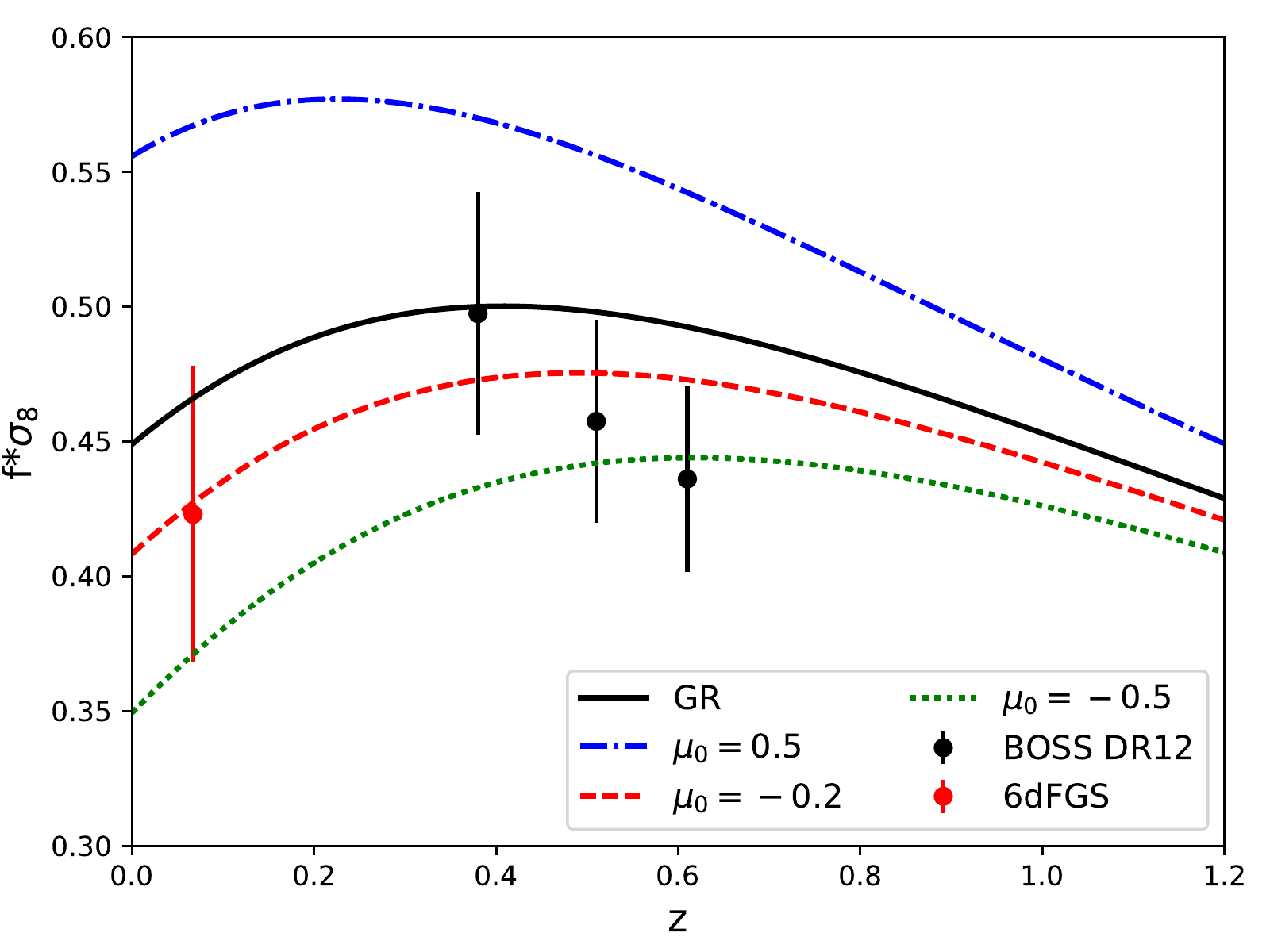}
\caption{The combination $f\sigma_8$ as a function of redshift $z$ for general relativity in black, $\mu_0 = 0.5,-0.2,-0.5$ respectively in blue dotted-dashed, red dashed and green dotted. The leftmost red point is the measurement from 6dFGS and the black points from BOSS DR12. The theory curves are uniformly shifted up or down when the power spectrum amplitude is varied.} 
\label{fig:rsd}
\end{center}
\end{figure}

Figure \ref{fig:rsd} shows the combination $f\sigma_8$ for different values of $\mu_0$, along with the measurements from BOSS DR12 (black circles) and 6dFGS (red circle). We see that a variation of $\mu_0$ by of order $\pm 0.5$ takes the predicted curve across the spread of the data, and hence we should expect an ultimate constraint of that order from this dataset. Higher $\mu_0$ leads to a higher amplitude of $f(z)\sigma_8(z)$, but also a change of the slope. Because $\mu_0$ is degenerate with $\sigma_8$, we therefore need other measurements such as the CMB in order to constrain the cosmology as part of constraining ($\Sigma_0$,$\mu_0$).

Constraints on modified gravity combining the measurements from BOSS DR12 and {\it Planck} have already been explored in Ref.~\cite{Mueller_2016}. The parametrisation used was similar to the present one ($G_L$ and $G_M$ in that paper correspond to our $\Sigma_0$ and $\mu_0$ respectively), but it adopted a different time parametrisation and did not include weak-lensing data.

\subsubsection{Results}

\begin{figure}[t]
\begin{center}
\includegraphics[scale=0.4]{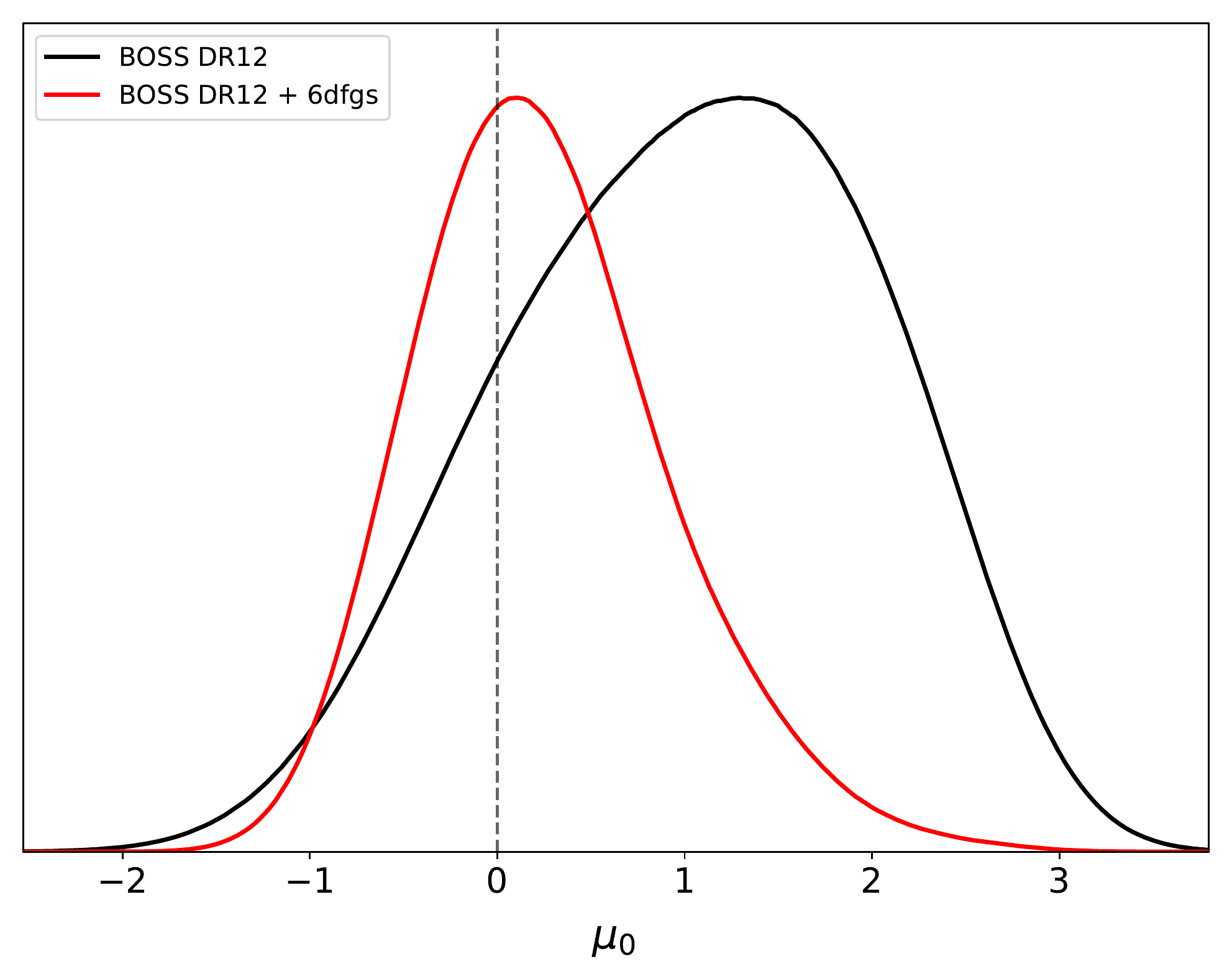}
\caption{The 1D posterior distribution of $\mu_0$, using BOSS DR12 data alone in black and combined with 6dFGS in red.}
\label{fig:mursd}
\end{center}
\end{figure}

As the RSD do not constrain $\Sigma_0$, we show in Fig.~\ref{fig:mursd} the 1D posterior distribution on $\mu_0$. The black line shows the result for BOSS DR12 alone, which mildly favors a positive $\mu_0$. Combining it with 6dFGS lowers the favored $\mu_0$, which can be understood from Fig.~\ref{fig:rsd} where the 6dFGS measurement sets the low-redshift value of $f\sigma_8$. In the case of the BOSS DR12+6dFGS, the peak of the 1D posterior distribution along with its uncertainty is
\begin{eqnarray}
\mu_0 & = &  0.0^{+0.6}_{-0.7}.
\end{eqnarray}
$\Lambda$CDM is consistent with these constraints. This constraints would correspond to the $G_M$ constraints in the `s = 3' case of Table 2 of Ref.~\cite{Mueller_2016}, although the latter also includes CMB data, yielding better constraints. 

\subsection{Cosmic microwave background}

\subsubsection{Observable and data set}

The anisotropies of the cosmic microwave background (CMB) temperature and polarisation, as well as providing the most precise estimate of the cosmological parameters, also constrain modified gravity, as extensively explored in Ref.~\cite{planck_demg_2015}. Several features arise in different cases of dark energy or modified gravity theories, as discussed in that paper. In this analysis, we will focus on the impact of modified gravity on the low multipoles of the CMB temperature and polarisation power spectra via the Integrated Sachs--Wolfe (ISW) effect and CMB lensing. 

\begin{figure}[t]
\begin{center}
\includegraphics[scale=0.5]{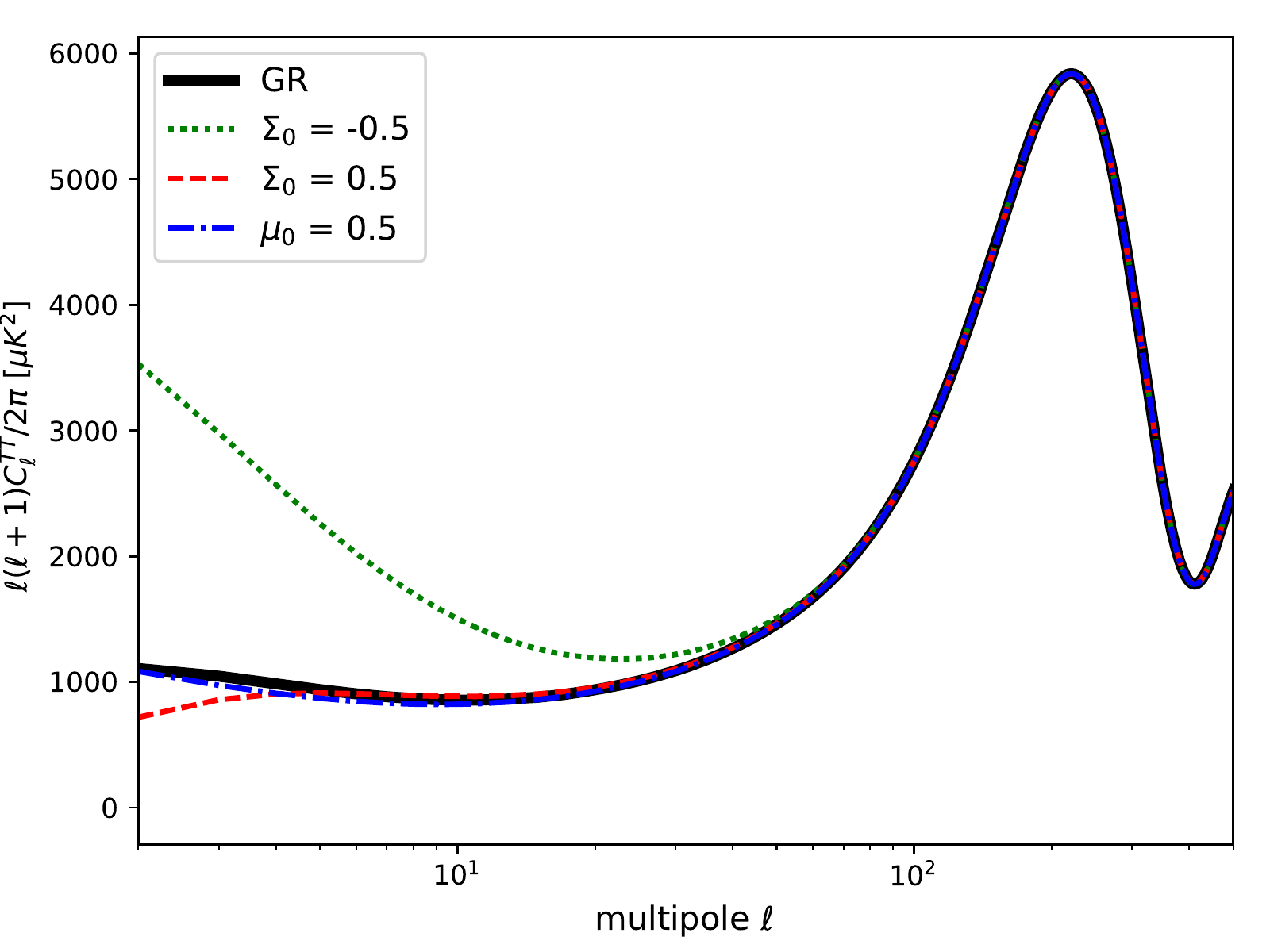}
\caption{Theoretical temperature correlations for different values of $\Sigma_0$ and $\mu_0$. The general relativity case is the black curve while $\Sigma_0 = -0.5$ corresponds to the red (dashed) curve and $\mu_0 = 0.5$ to the blue (dash-dotted) line.}
\label{fig:cmbttte}
\end{center}
\end{figure}

Figure \ref{fig:cmbttte} shows the effect of a non-zero $\mu_0$ or $\Sigma_0$ on the temperature power spectrum $C_{\ell}^{\rm TT}$. As the ISW effect arises from the time derivative of the combination \mbox{$\phi$ + $\psi$}, this observable is mostly sensitive to $\Sigma_0$, as we can see in the figure. At higher $\ell$ the curves overlap, with the modifications to gravity having no visible effect on the power spectra.
The variation induced by $\mu_0$ is small, so we expect the CMB to be weaker than RSD in constraining this parameter. By contrast, the curve with $\Sigma_0 = -0.5$ is easily excluded by current CMB observations, implying a constraint significantly stronger than this.

The {\it Planck} satellite was dedicated to the observation of the CMB temperature and polarisation anisotropies. 
In our analysis we use the TT likelihood for the high multipole part and the joint TT, TE, EE and BB likelihood for the low multipole part (which we will refer to as TT+lowP)  from the 2015 data release \cite{Planck_likelihood_2016}. We also use the {\it Planck} lensing measurements of the projected gravitational potential $\phi$.

\subsubsection{Results}

\begin{figure}[t]
\begin{center}
\includegraphics[scale=0.45]{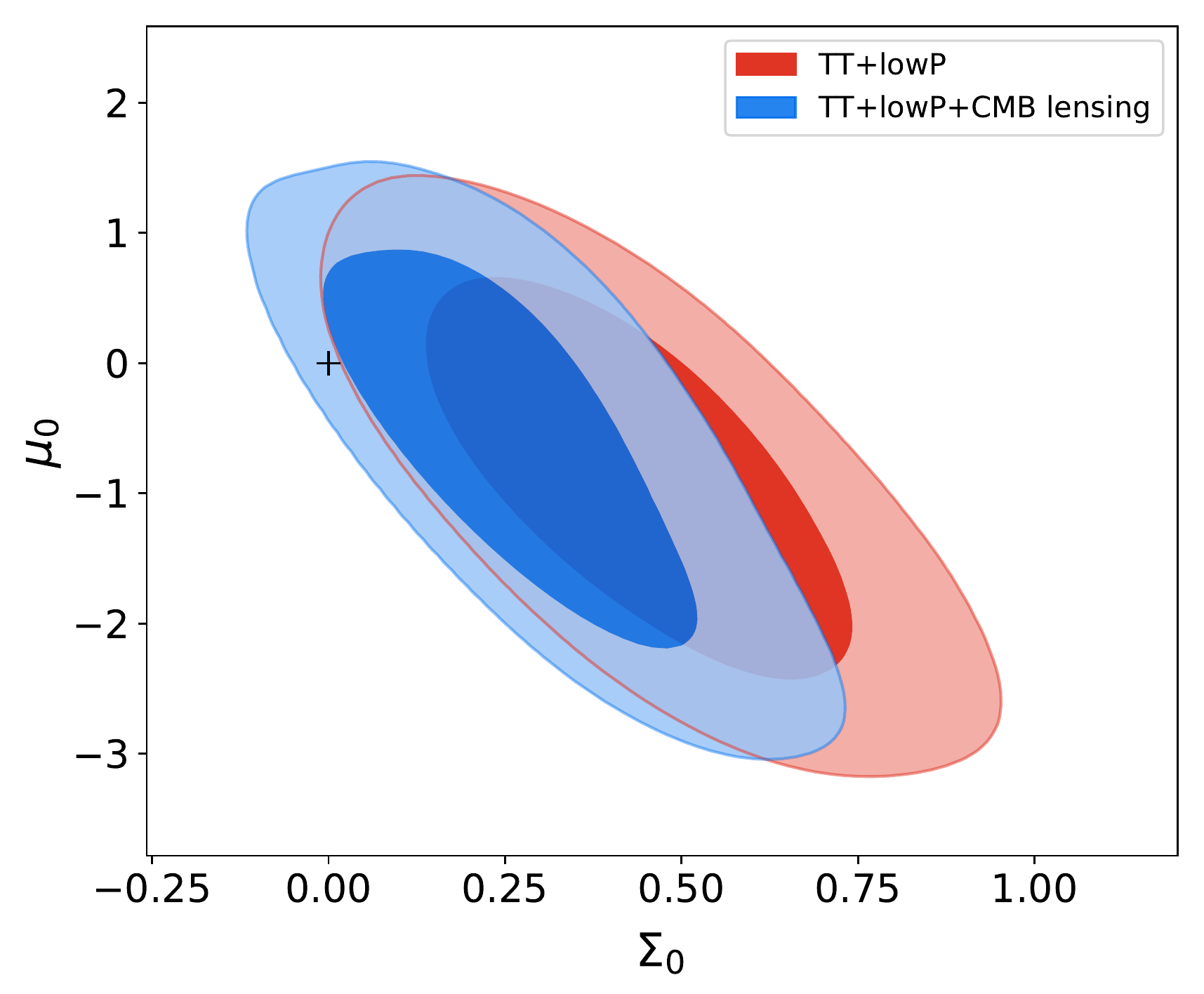}
\caption{68$\%$ and 95$\%$ confidence contours on $\Sigma_0$ and $\mu_0$, using temperature and polarisation \textit{Planck} likelihood in red, and the same adding the \textit{Planck} lensing likelihood in blue.}
\label{fig:planck}
\end{center}
\end{figure}

Using the {\it Planck} `lite' likelihood, we marginalize over the calibration amplitude $A_{\mathrm{P}}$, along with the cosmological parameters and ($\Sigma_0$,$\mu_0$) as listed in Table~\ref{tab:priors}. Figure \ref{fig:planck} shows the contours on ($\Sigma_0$,$\mu_0$) using only TT+lowP likelihoods in red and adding the CMB lensing in blue. As expected the constraining power on $\Sigma_0$ is higher than on $\mu_0$. Using TT+lowP alone leads to a slight tension with GR, vanishing when adding the CMB lensing which fixs the amplitude of the lensing $A_L$ closer to 1 and therefore favoring lower $\Sigma_0$ values. This tension has been seen and described by the {\it Planck} team in Ref.~\cite{planck_demg_2015}. 

For TT+lowP, we find the peak of the one-dimensional distributions and the uncertainties on ($\Sigma_0$,$\mu_0$) to be
\begin{eqnarray}
\Sigma_0 & = &  0.44_{-0.21}^{+0.18} \\
\mu_0 & = & -0.7_{-1.1}^{+0.9} .
\end{eqnarray}
On adding CMB lensing we find
\begin{eqnarray}
\Sigma_0 & = & 0.24^{+0.17}_{-0.16} \\
\mu_0 & = & -0.2^{+0.7}_{-1.3}.
\end{eqnarray}

\section{Combination of Probes to constrain modified gravity}
\label{sec4}

We have shown the constraints of the individual probes we consider in this analysis. This section focuses on the results from the combination of these datasets, here considered to be independent. 

\begin{figure}[t]
\begin{center}
\includegraphics[scale=0.5]{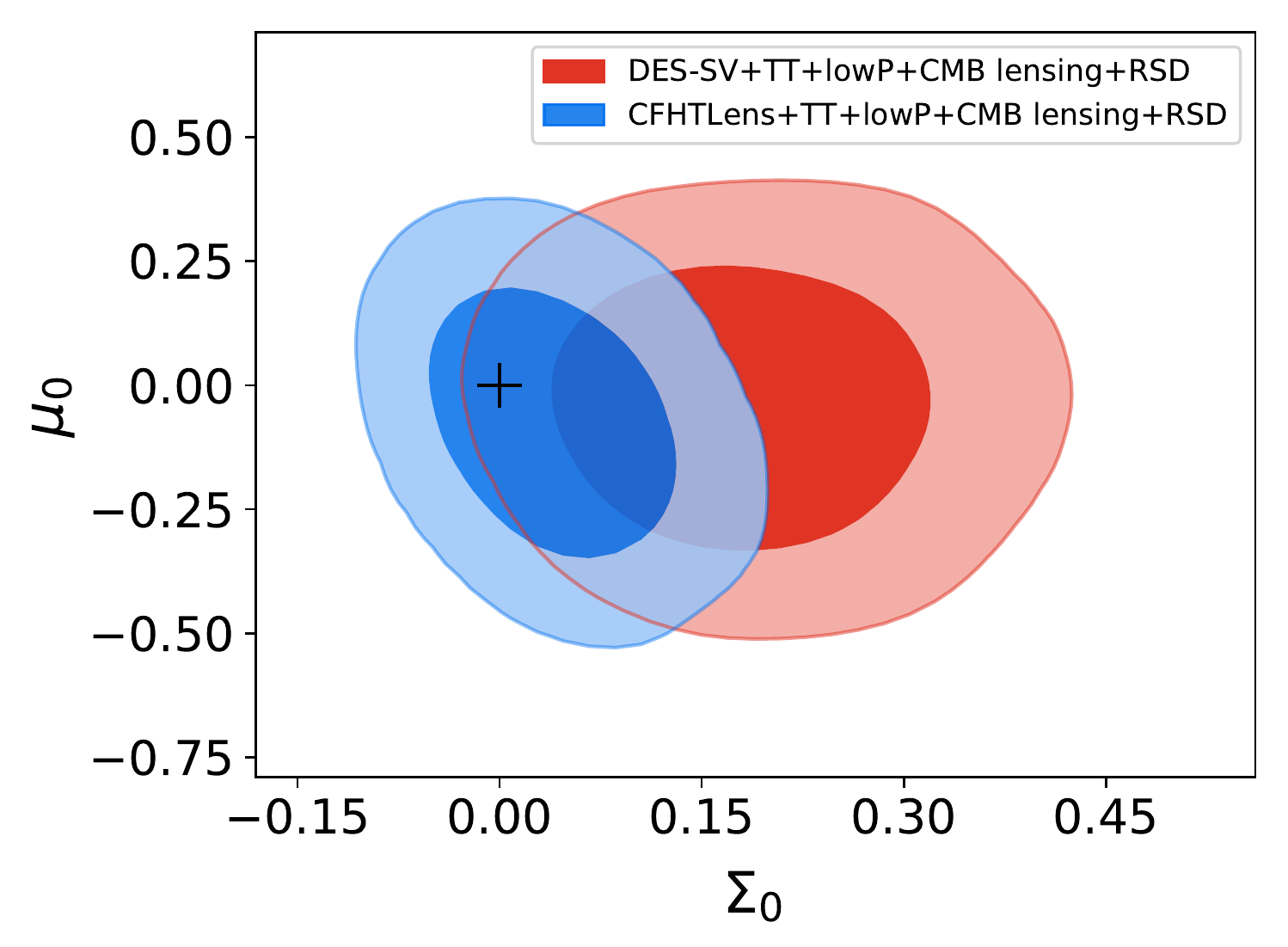}
\caption{68\% and 95\% confidence contours on $\Sigma_0$ and $\mu_0$ combining RSD data (BOSS DR12 + 6dFGS), CMB data (TT+lowP+CMB lensing from \textit{Planck}) and cosmic shear data (CFHTLenS in blue and DES-SV in red).}
\label{fig:comb}
\end{center}
\end{figure}

We combine the data from \textit{Planck} TT+lowP+CMB lensing, BOSS DR12 and cosmic shear data, varying the parameters in Table \ref{tab:priors}. We show this combination in Fig.~\ref{fig:comb} with DES-SV as cosmic shear data in red and CFHTLenS in blue. Using the DES-SV data gives wider constraints than from CFHTLenS as expected from its statistical power (see \textit{e.g.}\ the purple contours against the orange one in Fig.~10 of Ref.~\cite{dessv_2016}). We marginalise over the intrinsic alignment amplitude for the DES-SV data as well and find an amplitude similar to Fig.~8 of Ref.~\cite{dessv_2016}. A positive amplitude of the intrinsic alignment $A_{IA}$ is favored, leading to higher $\Sigma_0$. The blue contours are amongst the tightest constraints on ($\Sigma_0$,$\mu_0$) at present. We obtain:
\begin{eqnarray}
\Sigma_0 & = & 0.05_{-0.07}^{+0.05} \\
\mu_0 & = & -0.10^{+0.20}_{-0.16}  \,
\end{eqnarray}
GR is consistent with this combination of datasets. The constraints on $\Sigma_0$ are tighter than from earlier studies because of our use of the quasi-linear scales and of our combination of latest RSD and CMB measurements.  

Other time parametrisations such as $\Sigma = \Sigma_0 a^s$, or the introduction of a scale dependence have been proposed in the literature.The constraints on ($\Sigma$,$\mu$) using these alternative parametrisations would introduce additional parameters, therefore weakening the constraints. However these parametrisations would be interesting to study in the case of future surveys such as DES, LSST and EUCLID.

\section{Cosmic shear and modified gravity}

In this section, we forecast the expected constraints on ($\Sigma_0$,$\mu_0$) from future surveys, using the Fisher matrix technique. We use the Fisher module built into CosmoSIS. 

\subsection{Forecasts for DES Y5}
\label{sec:sys}

We now make forecasts on the constraints on $(\Sigma_0,\mu_0)$ for the complete 5 years of observation of DES (DES Y5) and a stage-IV like experiment such as LSST. We use a Fisher matrix, computing the expected uncertainties around a fiducial $\Lambda$CDM model. In our analysis, we marginalise over the five other cosmological parameters following Table~\ref{tab:priors}. 
We consider specifications as listed in Table~\ref{tab:fisher} with the sky coverage in $\deg^2$, the number of redshift bins, the number density of galaxies $n_{\mathrm{gal}}$ in $/\mathrm{arcmin}^2$, the intrinsic ellipticity standard deviation $\sigma_{\epsilon}$, and the standard deviation of the photo-$z$ estimation $\sigma(z)$ as a function of the redshift z. We ignore the effect of systematics for simplicity and fix the amplitude of the intrinsic alignment to unity. We further examine the effect of intrinsic alignments below. 

\begin{figure}[t]
\begin{center}
\includegraphics[scale=0.45]{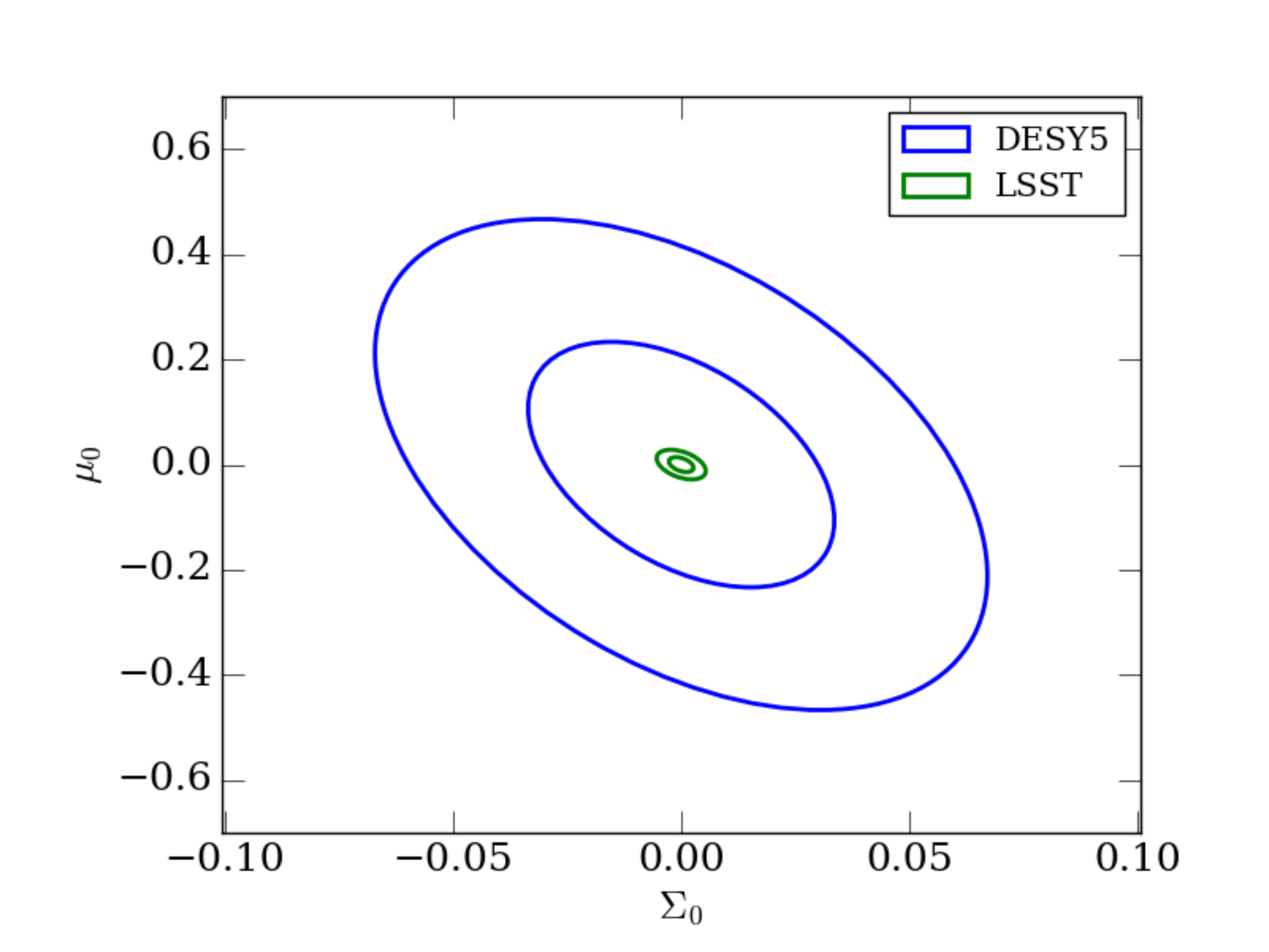}
\caption{Forecasted 68\% and 95\% confidence contours on ($\Sigma_0$,$\mu_0$) for a DES Y5-like like survey in blue and for LSST-like in green. }
\label{fig:mg_desy5}
\end{center}
\end{figure}

\begin{table}[t]
\centering
\begin{tabular}{|l|c |c|} 
 \hline
&  \textbf{DES Y5} & \textbf{LSST} \\
 \hline
Sky coverage ($\deg^2$) & 5000 & 18000 \\
Number redshift bins & 5 & 10 \\
$n_{\mathrm{gal}}$ ($/\mathrm{arcmin}^2$)& 10 & 55 \\ 
$\sigma_{\epsilon}$ per bin & 0.25 & 0.2 \\
$\sigma(z)$ & $0.05(1+z)$ &  $0.02(1+z)$	\\
\hline
\end{tabular}
\caption{Specifications for DES Y5 and for an LSST-like survey, used in our Fisher analysis.}
\label{tab:fisher}
\end{table}

Figure \ref{fig:mg_desy5} shows the forecasted constraints on ($\Sigma_0$,$\mu_0$) around their values in GR for the shear measurements from DES Y5 in blue and LSST in green.
We obtain a projected uncertainty on $\Sigma_0$ and $\mu_0$ of
\begin{eqnarray}
\sigma_{\Sigma_0} & = & 0.034, \; \\
\sigma_{\mu_0} & = &  0.23,
\end{eqnarray} for DES Y5, and
\begin{eqnarray}
\sigma_{\Sigma_0} & = & 0.0027, \; \\
\sigma_{\mu_0} & = &  0.014,
\end{eqnarray} for LSST. 

LSST gives standard deviations on ($\Sigma_0$,$\mu_0$) an order of magnitude smaller than those from DES Y5. The shear measurements from LSST seem therefore competitive with current RSD measurements in constraining $\mu_0$. 
This indicates that the increase in the number of tomographic bins, sky coverage, and galaxy density, as presented in Table \ref{tab:fisher}, are key to the expected improvement of the constraints on MG.
This analysis is optimistic but shows the power of cosmic shear from future weak lensing surveys to test laws of gravity. 

\subsection{Intrinsic Alignment}
\label{sec:IA}

In this subsection, we make a preliminary investigation of the impact of intrinsic alignments on MG constraints with a DES Y5 like survey. We refer the reader to Ref.~\cite{Krause_2016} for a comprehensive Fisher analysis of the impact of IA on cosmic shear surveys and dark energy constraints, in particular on equation of state parameters ($w_0$,$w_a$). In this part, we fix the value of $\tau$ to 0.08 as DES alone is not sensitive to $\tau$ but we can expect it to be well constrained by CMB measurements in any future analysis. 

\begin{figure}[t]
\begin{center}
\includegraphics[scale=0.45]{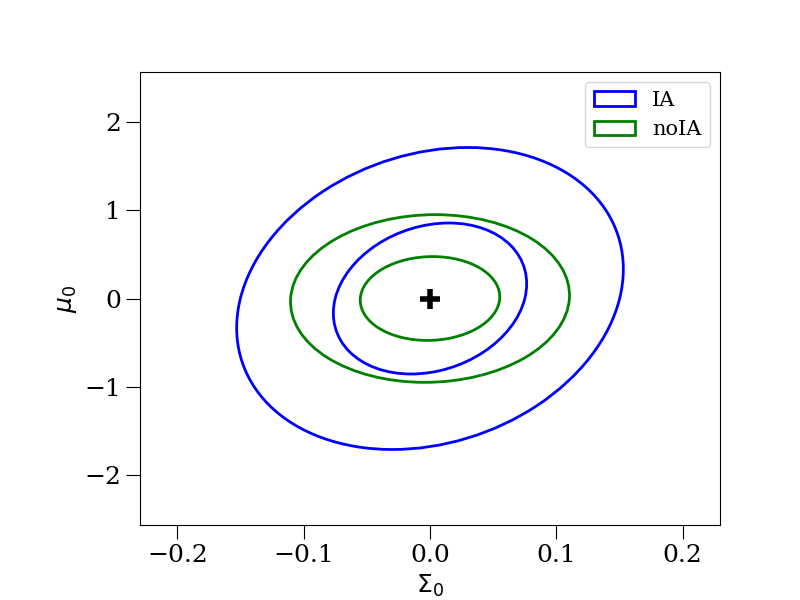}
\caption{Forecasted 68\% and 95\% confidence contours on ($\Sigma_0$,$\mu_0$) for a future DES Y5  like survey, ignoring intrinsic alignment in green and taking it into account in blue. }
\label{fig:mg_desy5b}
\end{center}
\end{figure}

It is expected that the non-linear IA model will be well constrained by DES Y5 in GR, even with a free amplitude per bin. So we will adopt this, which is the model leaving most freedom for the non-linear IA modelling.  Figure~\ref{fig:mg_desy5b} shows the constraints on ($\Sigma_0$,$\mu_0$) when marginalising over the amplitude of the non-linear intrinsic alignment model per redshift bin (so 5 parameters) in blue. In green, we don't take any intrinsic alignment in our modelling and therefore don't marginalise over the amplitude of IA. The latter predicts tighter constraints on ($\Sigma_0$,$\mu_0$), which can be understood as having less parameters to constrain. This shows ignoring the presence of IA in the data leads to underestimation of the uncertainties on the estimated MG parameters. 
\\
\section{Conclusions and discussion}
\label{sec5}

In this analysis, we have derived the constraints on time-dependent ($\Sigma_0$,$\mu_0$) parameters to test the laws of gravity on large cosmological scales, using three different probes of the evolution of large-scale structures: RSD measurements, the CMB temperature, polarisation anisotropies and lensing, and cosmic shear. We used the data from BOSS DR12, the \textit{Planck} satellite, and CFHTLenS survey. We find $\Sigma_0 = 0.05_{-0.07}^{+0.05} $ and $\mu_0 = -0.10^{+0.20}_{-0.16} $, indicating that these datasets are consistent with general relativity and substantially constrain possible deviations from it.

We are interested in the potential of cosmic shear in the framework of Modified Gravity. We  explored forecasts for the forthcoming DES Y5 and LSST data with a Fisher matrix forecast, finding $\sigma_{\Sigma_0} = 0.034$, and $\sigma_{\mu_0} = 0.23$ for DES Y5 and a further gain of an order of magnitude with LSST. These provide optimism that future surveys will have great constraining power on these parameters. Moreover cosmic shear is subject to systematics, the main ones being shear calibration bias, photometric redshift bias and intrinsic alignment. We show that the latter is the most important in the framework of MG, as ignoring it can lead to a bias in the estimated parameters. 

A significant limitation of this study on the theory side is the assumed scale independence of the MG parameters. This is enforced as present data lack the quality to meaningfully test variations, though it remains well motivated in the context of probing the ongoing viability of GR. The constraining power of future surveys will enable exploration of scale and time dependence of the deviations from general relativity.

\acknowledgements

A.F.\ and A.R.L.\ were supported by the Science and Technology Facilities Council [grant number ST/K006606/1], and A.R.L.\ in part by FCT (Portugal). DK acknowledges support from a European Research Council Advanced Grant FP7/291329. Research at Perimeter Institute is supported by the Government of Canada through Industry Canada and by the Province of Ontario through the Ministry of Research and Innovation. A.F.\ thanks Fergus Simpson and  Catherine Heymans for helpful discussions, and St\'ephane Ili\`c for discussions about parameter estimation and \textit{Planck} data.

\bibliographystyle{ieeetr}
\bibliography{paper_AFetal_2017}{}

\end{document}